\documentstyle[11pt,epsf]{article}
\input psfig
\def\beq{\begin{eqnarray}}
\def\eeq{\end{eqnarray}}
\def\bea{\begin{eqnarray*}}
\def\eea{\end{eqnarray*}}

\def\centeron#1#2{{\setbox0=\hbox{#1}\setbox1=\hbox{#2}\ifdim
\wd1>\wd0\kern.5\wd1\kern-.5\wd0\fi
\copy0\kern-.5\wd0\kern-.5\wd1\copy1\ifdim\wd0>\wd1
\kern.5\wd0\kern-.5\wd1\fi}}
\def\ltap{\;\centeron{\raise.35ex\hbox{$<$}}{\lower.65ex\hbox{$\sim$}}\;}
\def\gtap{\;\centeron{\raise.35ex\hbox{$>$}}{\lower.65ex\hbox{$\sim$}}\;}

\def\singleandthirdspaced{\baselineskip=\normalbaselineskip\multiply
    \baselineskip by 130\divide\baselineskip by 100}
\def\singlespaced{\baselineskip=\normalbaselineskip}

\newcommand{\newc}{\newcommand}
\newc{\qbar}{{\overline q}}
\newc{\Kahler}{K\"ahler }
\newc{\deltaGS}{\delta_{\rm GS}}
\begin{document}
\begin{titlepage}
\begin{flushright}
{\large hep-th/0405256 \\ SCIPP-2004/13\\
}
\end{flushright}

\vskip 1.2cm

\begin{center}

{\LARGE\bf Is There a Peccei-Quinn Phase Transition?}

\vskip 1.4cm

{\large  M. Dine$^a$ and A. Anisimov$^{b,c}$}
\\
\vskip 0.4cm

{\it $^a$Santa Cruz Institute for Particle Physics,
    Santa Cruz CA 95064  } \\
{\it $^b$Max-Planck Institute fuer Physik, Munich}\\
{\it $^c$Institute of Theoretical and Experimental Physics, Moscow} \\
\vskip 4pt

\vskip 1.5cm

\begin{abstract}

The nature of axion cosmology is usually
said to depend on whether
the Peccei-Quinn (PQ) symmetry breaks before or after inflation.
The PQ symmetry itself is believed to be an accident, so there
is not necessarily a symmetry during inflation at all.
We explore these issues in some simple models, which provide
examples of symmetry breaking before and after inflation,
or in which there is no symmetry during inflation and no
phase transition at all.
One effect of these observations is to relax the constraints
from isocurvature fluctuations due to the axion during inflation.
We also observe new possibilities for evading the constraints due to
cosmic strings and domain walls, but they seem less generic.

\end{abstract}

\end{center}

\vskip 1.0 cm

\end{titlepage}
\setcounter{footnote}{0} \setcounter{page}{2}
\setcounter{section}{0} \setcounter{subsection}{0}
\setcounter{subsubsection}{0}

\singleandthirdspaced

\section{Introduction}

Traditional discussions of the axion in cosmology begin with the
assumption that the Peccei-Quinn symmetry
is exact at all times in cosmic history, and
that the universe was in thermal equilibrium at the
time of the Peccei-Quinn (PQ) phase transition and thereafter.  This
leads in a rather straightforward way to an upper limit on the
axion decay constant (and a lower limit on the axion mass)\cite{axionlimits}.
In non-supersymmetric models, such a scenario may seem plausible.
However, the low axion scale is likely to require fine tuning, 
or, more plausibly, some new strong dynamics which might modify 
the naive picture.

But in many contexts these assumptions do not seem reasonable.  If 
nature is supersymmetric, they are unlikely to hold.  
Many authors have discussed the cosmology of a possible light scalar field
(called the saxion\cite{saxion}) which accompanies the axion\cite{saxioncosmology}.
In \cite{bdg}, it was stressed that this particle is necessarily
a modulus or approximate modulus.  It was
noted that the coherent motion of this modulus is likely to 
dominate the energy density of the universe long before the QCD 
phase transition and that string theory suggests that there might, in 
fact, be many such moduli.
The cosmological problems
associated with these additional moduli are parameterically much
more severe than those associated with the axion, so it does
not make sense to discuss axion cosmology without
first resolving these.  Some
proposed solutions to these problems significantly relax the usual
axion limit.  In particular, decays of the moduli dilute the
axion, permitting decay constants as large as $10^{15}$ GeV.

The discussion of \cite{bdg} assumed that there was a stage of
inflation before PQ symmetry breaking.  There is a widely
held prejudice that this is unlikely.  This prejudice is based on the
plausible assumption that the PQ transition occurs when
the temperature of the universe is comparable to the PQ scale.
If the scale is, say, $10^{11}$ GeV, then this might be expected
to be well after inflation.

Again, though, in the context of supersymmetry, this reasoning is
suspect.  Our arguments above suggests that the universe might not be
radiation dominated until rather late in its history.  In
supersymmetry, in addition, the gravitino problem indicates that
after inflation the universe did not reheat to a temperature as
high as the PQ scale.  The aim of this paper is to enumerate some of
the possible PQ phase transitions in supersymmetric theories and their
implications for axion cosmology.   In the next section, we
will review a class of field theory models (sometimes called
``flatton models") in which the
PQ symmetry arises as an accident of discrete symmetries
of the low energy
effective theory.  In these models, we will see that depending on the
transformation properties of the inflaton(s) and the signs
of certain terms in the effective action,
the PQ transition can\footnote{This point discussed below
that the sign of an effective mass term can take either value
during inflation, and determines the nature of the transition,
 has been made by a number of authors
\cite{Linde:1991km,lythsign}.  That this would
be relevant to the question of isocurvature perturbations
was already discussed in Linde's paper.  More recently, some of the effects of this
term during inflation have been discussed in \cite{Dimopoulos:2003ii,Chun:2004gx},
with particular attention to the possibility that the axion
itself is the source of the observed fluctuations.}
occur readily {\it either} during or after inflation, or there
may be no transition at all.  If the transition
occurs before (or during) inflation, there is a potential problem
with isocurvature fluctuations\cite{axioniso}.   As we will discuss, the decay constant
of the axion initially takes a value generally quite different than its
value at late times.  This has possible
implications for isocurvature fluctuations (some aspects of this problem
were discussed in \cite{Linde:1991km,lythiso}).

Potentially more important is the fact that there is quite possibly no phase transition
at all.  More precisely, during inflation, there may be no Peccei-Quinn symmetry.  Some
time after
inflation ends, there would be an approximate symmetry, which would get better and better
with the passage of time.  By the time the symmetry can be thought of as a good
symmetry, it is non-linearly realized, i.e. it is already spontaneously broken.

In section three, we will turn to the case where the phase transition occurs
after inflation.  The usual picture of this transition is that in different
domains, the axion field is essentially a random variable.
This leads to the formation of axion cosmic
strings and domain walls, as well as an
alignment problem.  But we will see that, generically, there
is a period after the PQ phase transition in which the
PQ symmetry is badly broken.  Indeed, for a rather long time,
one can have $m_a >> H$.  As a result, the axion can be
driven to a particular point.  One might then hope to avoid
domain walls and strings, and perhaps solve the alignment problem,
as in the first case, through dilution.  We will see, however, that while the
symmetry is broken, the system is typically driven not to one but
to several points.  This leads, again, to the domain wall
problem.  We will discuss these issues, and a subclass
of models which might evade these problems. 

In section four, we briefly discuss string theory models.  It is
logically possible that string models could reproduce any of the
phenomena of the field theory models.  But our current understanding
of PQ symmetry breaking in M theory is too rudimentary to
allow us to say much more.  In section
5, we conclude with an overview of recent results on axion cosmology,
and assess the significance of experimental limits on axion production.
In particular, we stress that without further knowledge of microscopic physics,
it is difficult to place cosmological constraints on axions.

\section{Field Theory Models and Their Possible Cosmologies}

In four dimensional field theories, one way to obtain
PQ symmetries is as accidental consequences of discrete
symmetries\cite{Lazarides:bj}.  In supersymmetric theories, this idea is readily
implemented\cite{wittensymposium}.  Models of this type
suggest a very specific
picture for the PQ phase transition

\subsection{A Simple Model}

As a simple example, consider a theory with fields $S$ and
$S^\prime$ (one can consider models with a single scalar field,
but in this case the PQ symmetry is an $R$ symmetry, and
it is difficult to reconcile with the various constraints on
soft-breaking terms).  Suppose that the model possesses a $Z_N$
symmetry under which \beq S \rightarrow \alpha S ~~~~~ S^{\prime}
\rightarrow \alpha^{N-n} S^\prime. \eeq
Then the leading allowed
terms in the superpotential are
 \beq W={1 \over M^{n-2}} S^n
S^\prime + {1 \over M^{N-3}}S^N + \dots 
\label{ssuperpotential}
\eeq
Here $M$ denotes the (reduced) Planck scale and the dots indicate
terms with more than one power of $S^\prime$ as well as higher
powers of $S$.  (Our arguments can be reworked if scales other than
the Planck scale are important).

Now we expect that the soft-breaking terms include:
 \beq
 {\cal L}_{soft}= -m_{3/2}^2 \vert S^2 \vert + m_{3/2}^2 \vert S^\prime \vert^2
 + A {m_{3/2}\over M^{n-2}}S^n S^\prime +  \dots
 \label{softterms}
 \eeq
The potential has an approximate PQ symmetry under which
\beq
S \rightarrow S e^{i \omega}. \eeq
The leading
symmetry-breaking operators are the terms in the superpotential
indicated above of the form ${1 \over M^{N-3}} S^{N}$ as well as
soft breaking terms,
\beq {m_{3/2} \over M^{N-3}}S^N.
\eeq

The potential has a minimum with
\beq 
S= (m_{3/2} M^{n-2})^{1 \over
n-1}.
\eeq
This spontaneously breaks the PQ symmetry,
giving rise to an axion.
Note
that for $n=3$, $S=f_A \sim 10^{11}$.  For $N \ge 11$, the explicit symmetry breaking
is
small enough that this axion can solve the strong CP problem.   More elegant models
can be constructed with products of smaller discrete symmetries, such as frequently
arise in string theory.

\subsection{The Crucial Question:  Does the Inflaton Transform Under
the Discrete Symmetries?}

Now consider the early universe.  In the supersymmetric case, we suppose
that we have an inflaton, $I$, with non-vanishing $F$-component. The
Hubble constant during inflation is then: \beq H^2 \approx ({F_I^2 \over
M^2}, {W^2 \over M^4}). \eeq
For definiteness, we will assume that the Hubble constant during inflation is
at least as large as the weak scale, and that the expectation value of
the inflaton is of order the Planck scale, $M$.  It is straightforward
to modify our discussion for other possibilities.
After inflation, we will assume that the inflaton undergoes a period of
oscillation, during which the universe is matter dominated.
Now there are several possible phase structures, depending on the transformation properties of
the inflaton under the discrete symmetry.  The simplest case to consider is
that the inflaton is neutral.  Then the effect of the
couplings of the inflaton to the field $S$ is mainly to give
supersymmetry breaking of order $H^2$\cite{drt}.  The effective potential
now includes soft breaking terms of the form of eqn. \ref{softterms}, but
with $m_{3/2}$ replaced by $H$:
\beq
 {\cal L}_{soft}= (a ~H^2-m_{3/2}^2) \vert S^2 \vert + (b~H^2+m_{3/2}^2) \vert S^\prime \vert^2
 + \dots
 \label{learly}
\eeq

Depending on the sign of these terms the transition
between the broken and unbroken phase occurs in quite different
epochs of the universe.  If, during
inflation, the constant $a$ is negative, the phase transition will
occur at the Hubble-scale of inflation, $H=H_{inf}$.  In many
models of inflation, this is long before $H=m_{3/2}$.
In theories of hybrid inflation, along the lines of \cite{guthrandall},
the scale can be comparable to $m_{3/2}$.
The
axion field will be homogeneous after inflation, and there will be no issues of cosmic
strings or domain walls.  One will have a problem of alignment.
If $a$ is positive the transition occurs after inflation, once $H\sim m_{3/2}$
(in the case of hybrid inflation, it occurs after
inflation in the era when the additional
modulus is settling to its minimum).

It is quite likely that the inflaton itself transforms
under the discrete symmetry.  In this case, there is a richer
set of possible behaviors.  First, if
if $I \sim M$,
then $S$, and
$S$ and/or $S^\prime$ will remain light only if they are protected
by symmetries unbroken by the inflaton.   If they are both heavy, the transition
can occur only well after inflation has ended.  In other words,
if the PQ symmetry is to occur during inflation, it is necessary
that the inflaton be neutral at least under a subset of the discrete
symmetries.  Even if $S$, $S^\prime$
are light, their couplings to the inflaton
will typically ``break" the discrete symmetry, i.e. there will
be no approximate PQ symmetry during this
early era.  We will explore some of the possible phenomena in this case
in the next section.

\subsection{PQ Breaking During Inflation}

Even if the PQ symmetry breaking occurs during inflation, the axion cosmology
is different than
usually assumed.  Consider, first, the case
where the inflaton is neutral under all of the discrete
symmetries.  Then there is
still a Peccei-Quinn symmetry
during inflation, but the axion decay constant during inflation is significantly
larger than at later times.   This already weakens the constraints coming from the absence of
isocurvature fluctuations in the spectrum of the CMBR, permitting rather large values
of $H_{inf}$.  
If the mass of $S$ is negative during inflation while that of $S^\prime$ is positive,
then
\beq
f_a \sim S= (H_{inf} M^{n-2})^{1 \over
n-1}.
\label{largervalue}
\eeq
If both $S$ and $S^\prime$ have negative masses, $f_a$ will be substantially larger.
Indeed, if we only keep the operator of eqn. \ref{ssuperpotential}, then the $S^\prime$
vev can run off to $\infty$.  So the precise value of $f_a$ depends on the details of
the model, and it can be far larger than even suggested by the equation \ref{largervalue}
above. For example, if $n=5$, so that $f_a \sim 10^{15}$ today, it will be closer to
$10^{17}$ GeV during inflation if the $S$ mass is negative and the $S^\prime$ mass
positive.  If the $S^\prime$, mass is negative, then the value of $f_a$ depends on the
model.  In the $Z_{11}$ example, the leading operator which stabilizes the $S^\prime$
vev is $S^{\prime ~11}$, giving $f_a \sim 10^{17.5}$.

If the inflaton transforms under some of the discrete symmetry, as noted above, it is
necessary that enough symmetry survive that at least one of the fields $S$ or $S^\prime$
does not acquire a large mass (here we are assuming $I\sim M$,
so that operators involving arbitrarily large powers of
$I$ are not suppressed.  For example, $S^\prime$ might acquire a large mass,
but the leading operator giving rise to a superpotential for $S$
might have the form $I^n S^{m+3} \over M^{m+n}$.  
In this case, assuming $I \approx M$, the $S$ vev is of order
\beq
S \approx (M^{m} H_{inf})^{1 \over m+1}.
\eeq
This can be quite different, again, then the value today.  

But what is striking in this case is that there may be additional effects
which explicitly break the Peccei-Quinn symmetry.  For example, soft
breaking terms of the form
\beq
V_{soft}= H_{inf} S^{m+3}
\eeq
break the Peccei-Quinn symmetry, giving rise to a mass for the axion -- like
that of the real part of $S$ -- of order $H_{inf}$.  In other words, while we
are speaking here of PQ symmetry breaking during inflation, there is really
no sense in which there is a PQ symmetry at all!  Instead, after inflation ends,
the explicit breaking of the symmetry decreases with time.  There is no real
phase transition.

To summarize, the PQ transition could well occur during inflation.  There may
be a sense in which there is an approximate PQ symmetry at this time, or there
may not.  What are the implications of these various possibilities?

If the PQ transition occurs during inflation and the axion field
is light, it is well known that there are significant constraints
from the lack of observed isocurvature fluctuations.
In order that isocurvature fluctuations
be small enough to be consistent with recent CMBR measurements, one requires at least
\beq
{\delta \theta \over \theta} < 10^{-5}.
\eeq
As has recently been stressed by \cite{thomasetal}, observation
of gravitational waves from inflation would place $H \ge 10^{13}$,
and significantly constrain axion models.
With much lower scales of inflation, as can arise in hybrid inflation,
isocurvature perturbations are not a serious issue.

To assess the constraints on $H$ and $f_a$ during inflation,
recall that a light axion 
generates isocurvature fluctuations with the 
power spectrum of a massless field 
in deSitter spacetime
\beq
{\mathcal P}_a(k)={k^3\over 2\pi^2}|a_k|^2=\left({H_{inf}\over 2\pi}\right)^2,
\eeq
thus yielding for $\theta_a=a/f_a$ 
\beq
{\mathcal P}_{\theta_a}(k)=\left({H_{inf}\over 2\pi f_a}\right)^2.
\label{powspect}
\eeq
The expression (\ref{powspect}) is correct provided that the axion
mass is small compared to $H_{inf}$ during inflation. 

While during the radiation dominated period isocurvature fluctuations do not
generate any curvature perturbations, during the matter dominated period they
do, due to the
Sachs-Wolfe effect: ${\mathcal R}_a=
(1/3)S_a$\footnote{Here $S_a$ is the entropy fluctuation, generated by 
axion field $S_a=(\delta\rho_a/\rho_a)=2(\delta\theta_a/\theta_0)=
H_{inf}/(\theta_0\pi f_a)$, where 
$\theta_0$ is the initial value of misalignment angle.}
(see, for example, \cite{lyth_riotto}). 
For  scales larger than horizon size at matter-radiation equality 
(i.e. for low multipoles, $l<200$) the cmb temperature anisotropy 
coming from isocurvature fluctuations is a combination of the Sachs-Wolfe 
contribution and a contribution generated by entropy fluctuations 
\cite{lyth_liddle} 
\beq
\left({\delta T\over T}\right)_{\rm iso}=-{1\over 5}{\mathcal R}-{1\over 3}S=
-{6\over 15}\left({H_{inf}\over \pi f_a\theta_0}\right).
\eeq 
>From the recent WMAP data the limit on CDM isocurvature fraction in  
the CMBR spectrum at low multipoles is \cite{crotty}
\beq
\left({\delta T\over T}\right)_{\rm iso}^2<{\alpha\over 1-\alpha}
\left({\delta T\over T}\right)_{\rm ad}^2, 
\eeq
with $\alpha=0.31$. Taking all of the above into account\footnote{For
$(\delta T/ T)_{\rm ad}$ we took for simplicity COBE result  
$(\delta T/T)|_{10^o}=1.1 \times 10^{-5}$.}, we can put a 
constraint on the ratio of $H_{inf}$ to $f_a$
\beq
{H_{inf}\over f_a}<5.8\times 10^{-5} \theta_0\Omega_a^{-1},
\label{bound}
\eeq
where $\Omega_a$ is the axion fraction in the total matter density in the 
universe.
If we assume that $\theta_0\sim \Omega_a^{-1}\sim 1$, $H_{inf}\sim 
10^{13}{\rm GeV}$, this constrains $f_a$ to be larger then $10^{17}{\rm GeV}$
during inflation, so that isocurvature fluctuations may be at a barely
acceptable level.

A much
more important suppression of axion isocurvature fluctuations may come 
from the fact that during inflation the axion may acquire a large mass  
due to its coupling to the inflaton. For example, there could be
a superpotential term
\beq
W={1 \over M_p^{m-1}} S^{m+1} I.
\eeq 
Taking $S=S_o\exp(ia/S_0)$ with $S_o\sim (H_{inf}M^n)^{{1\over n+1}}$ gives rise to a mass for the axion 
\beq
m_a^2\sim H^2_{inf}\left({H_{inf}\over M}\right)^{{m-1\over n+1}-1}.
\eeq
For $m<n+2$ axion mass is parametrically larger then $H_{inf}$.

If this occurs, the power spectrum of axion 
fluctuations is no longer flat but rather
strongly suppressed on large 
scales by a factor of ${H_{inf}^2\over m_a^2}({k\over aH})^3$, thus giving no 
contribution to the observed CMBR.

%

\section{PQ Breaking After Inflation}

If the constant $a$ is positive, or if the superpotential is such
that it gives a large mass to $S$ and $S^\prime$,
the phase transition will not occur until $H\approx m_{3/2}$ or later.
This may be long after inflation.   The usual assumption in this case
is that the PQ field is a random variable.  This leads
to the production of domain walls and cosmic strings,
both of which are potentially problematic\cite{shellard}.  As discussed
in \cite{lythiso}, in this situation
there is likely to be a period
of thermal inflation (see, for example, \cite{lythsign}). But we will see that, as in the
previous section, the behavior of the system is very sensitive
to the transformation properties of the inflaton
under the discrete symmetry responsible for the PQ symmetry.  For a broad
range of possible transformation laws, the PQ symmetry
is explicitly, and significantly, broken.  The axion
acquires a mass which, for a significant period, is large compared
to the Hubble parameter.    

As a warmup, consider first the case that $I$ is neutral under the discrete
symmetries.  Then the question is just the sign of the constant $a$ in eqn. \ref{learly}.
If $a$ is positive, the transition occurs after inflation, and completes once $H$ is
somewhat less than $m_{3/2}$.  In this case, we have the usual problems of
misalignment, cosmic strings and domain walls.

Now suppose that $I$ does transform under the discrete symmetry.  Clearly
there are now myriad possibilities.  We will discuss a few for purposes of illustration.
Consider a superpotential which includes
inflaton couplings to the fields $S$ and $S^\prime$:
\beq
W = {S^\prime S^{n+2} \over M^n}+ I^2 { S^{m+1} \over M^m}.
\label{wchi}
\eeq
Our working assumption will be that $I \sim M$ immediately
after inflation.  (In particular, if $I$ transforms under the discrete
symmetry; after inflation, it has a Planck scale vev, but this decreases
to zero over time, restoring the symmetry).  

This model, for $I=0$, has an approximate PQ symmetry:
\beq
S \rightarrow e^{i\alpha} S ~~~~~S^\prime \rightarrow e^{-i(n+2)\alpha}S^\prime
\eeq
As usual, we will assume that, as a result of discrete symmetries, the leading
symmetry-violating terms (for $I=0$) are of very high dimension.

We will also assume that there is a soft breaking term:
\beq
V_{soft} = m_{3/2} I^2  {S^{m+1} \over M^n}+ \dots.
\eeq
(where the dots denote
other soft breaking terms which are either highly
suppressed or do not break the PQ symmetry).
Note that $I^2$ will average, in general, to a complex number;
had we taken a linear term, it would average to zero.

We will suppose that $m < n+2$.  Also we assume that, immediately at the end of inflation,
$I \sim M$.  So the value of $S$ (and $S^\prime$), immediately after inflation,
is much less than its value at late times.  Initially, the first term in the 
superpotential of eqn. \ref{wchi} dominates.  It respects its own global symmetry,
which is broken by the second term.  As time evolves, however, $I^2$ decreases
(like $H^2 e^{-\Gamma_I t}$) and the relative importance of
the second term in eqn. \ref{wchi} grows.
Particularly interesting is the time when the two terms are comparable.
At this time:
\beq
\vert S^{n+1}\vert \approx  M^{n} m_{3/2} ~~~~~ 
\vert S^\prime \vert \approx {\vert S^{*n+2} \vert 
\over m_{3/2} M^n}  ~~~~~~  <I^2>  \approx {m_{3/2} M^{m} \over
S^{m-1}}e^{i\delta}. 
\eeq
This leads to a potential for the phase of $S$, $S =  S_o e^{i a/ S_o}$
\beq
V(a) \approx m_{3/2}^2 S^2 \cos((m+1)a/S + \delta)
\eeq
This corresponds to an axion mass of order $m_{3/2}^2$, at a time when
the Hubble constant is much smaller.

We have studied the behavior of a number of such systems numerically.
We find that the modulus of the field does indeed track the minimum of the
potential, and that the axion also settles into its minimum, damping at a rate
somewhat slower than $1/t$.

As it stands, however, the model generates domain walls.
The origin of the
problem is the $m+1$ within the cosine.  This arises because, in the approximation
in which keep just these two terms in the superpotential, the model has a 
$Z_{m+1}$ symmetry, which is spontaneously broken by the expectation value of $S$.
So while the axion has mass which becomes large compared to the Hubble constant, it
has $m+1$ minima, and thus domains form in this model.  Were it not for the domains,
the axion field would be homogeneous (it is homogeneous on the Hubble scale), and domain
walls would not form at the QCD phase transition.

This problem would be avoided if
\begin{enumerate}
\item  There are additional fields involved in inflation (or the inflaton is a linear
combination of fields with different $Z_N$ charges).  In this case, there could
be terms like
\beq
{I^{\prime 2} S^\prime S^{m} \over M^m}.
\eeq
These give a comparable contribution to the axion potential, with a different periodicity.
\item  There could be terms of slightly higher dimension.  These lift the degeneracy between
the states.  The domain walls tend, then, to collapse.  However, at best, only one additional
power of $S$ ($S^\prime$) permits domain wall collapse within a Hubble time.
\end{enumerate}

In the first case, the discrete symmetry is explictly broken, and there is no
corresponding degeneracy among states.  However, there is still a potential
problem.  These complicated potentials may well have several local minima, and it is possible
that in some regions, the system will become trapped in these.  Whether or not
this occurs requires a painstaking, case by case analysis, which we have not
performed.  However, the general lesson seems to be that despite the large breaking
of the PQ symmetry, for most choices of discrete charge assignments, there remains
a domain wall problem.

This analysis indicates that, quite generally, the PQ symmetry is
badly broken immediately after inflation.  It also makes clear that the implications
of this breaking for the formation of topological defects depends on the details 
of the underlying theory. It appears that, if the Peccei-Quinn symmetry is broken during
inflation (either spontaneously, explicitly or both), the usual constraints from
isocurvature fluctuations are ameliorated or even eliminated.  In the case that the 
transition occurs later, this breaking can provide a solution of the domain wall problem,
but only in rather special circumstances.

\section{String Models}

Axions arise very naturally in string theory.  They can often be
thought of as accidental consequences of higher dimensional gauge
symmetries associated with antisymmetric tensor fields, but this
is not always the case.  These symmetries are
good of various perturbative expansions. theory.
But, in general, the size of non-perturbative violations
is not known.  In particular, while one can use discrete shift symmetries
to place constraints on the size of superpotential terms which break
the symmetry, there is, a priori, a wide variety of possible Kahler
potential corrections, and their parameteric dependence on the coupling
constants is not known\cite{bdaxionsize}.

In weakly coupled string phenomenology, the string and Planck scale are not
significantly different.  Then the axion decay constant is large,
of order $\alpha_{gut} \times M$\cite{choietal}.  The axion field is part of
a supermultiplet with other moduli.  The potential for the real
part of the modulus, $\phi$, is assumed to take the form:
\beq
V=m_{3/2}^2 f(\phi/M).
\eeq
The behavior of the imaginary part is more complicated and highly
model-dependent.  It has been discussed in
\cite{bdg}.

At early times, one expects a structure:
\beq
V=H^2 M^2 g(\phi/M).
\eeq
The function $g$ is in general unrelated to the function $f$; its minima, if any,
are not expected to lie near those of $f$.  As a result, during inflation, couplings
can easily differ by order one from their values now.  The axion decay constant
can also differ by an amount of order one from its value now; in some scenarios, it might
differ by several orders of magnitude.  The axion potential can easily differ by
orders of magnitude from its current value.

We have little general understanding of the non-perturbative breaking of PQ symmetry
string theory.  For example, we don't know if there might be Kahler potential effects of order
$e^{-1/g}$, or whether all effects are of order $e^{-a/g^2}$.  So we can only speculate.  We
might imagine that the possibilities are similar to those we described in effective
field theories with discrete symmetries.  For example, it is possible that the couplings
were much larger during inflation, and so the explicit breaking of the PQ
symmetry was large.

The recent suggestion that all moduli might be fixed in flux compactifications\cite{kklt}
provides another interesting framework in which one might hope to address these questions.
But at present, it is not clear how axions would arise in this framework.

Axions might also arise as accidents of discrete
symmetries, as in our field theory models.  Because of all of these uncertainties,
the only sensible approach at the moment would seem to be
to view the field theory models as
indicating the range
of possibilities.  

\section{Conclusions}         

The main lesson of this work is that the traditional analyses of axion cosmology are not
robust.  Already, in \cite{bdg}, it was stressed
that in supersymmetric models, inevitably the
constraints on the axion decay constant from
axion misalignment are altered.
In this paper, we have seen that the underlying phenomena
which give rise to the PQ symmetry, as well as the mechanism of inflation
strongly affect the PQ phase transition. This, we have seen, has implications
for inflationary fluctuations and the formation of defects after
inflation.
The PQ symmetry is most likely an accidental
consequence of some deeper, underlying symmetry, perhaps a gauged discrete symmetry
or a higher dimensional continuous gauge invariance.  In either case, the fate of the
axion in the early universe depends on how the inflaton (or whatever drives inflation)
influences this accident.  If the inflaton transforms, for example, under the 
discrete symmetry, we have seen that there are a host of possibilities.

It is perhaps worth closing
by summarizing the possible axion cosmologies we have enumerated here:
\begin{enumerate}
\item  The PQ symmetry may already be broken during inflation. This breaking
may be spontaneous, explicit, or both.  In all cases, the constraints
from isocurvature fluctuations are altered, and in the explicit
case, they may be eliminated altogether. 
\item  The PQ transition may occur after inflation.  The usual picture, with formation
of cosmic strings and domain walls, may be correct.  Alternatively, for an epoch after
inflation ends, when the universe is dominated by the oscillating inflaton and/or moduli,
the PQ symmetry may be explicitly -- and badly -- broken by inflaton dynamics.
In this case, the axion may be aligned.  Depending on the details of the underlying
theory, this may eliminate the problems of domain walls, or it may not.  
One might imagine that states without such problems might be selected by rather mild
anthropic considerations.  
\end{enumerate}

All of this suggests that the breaking of the 
PQ symmetry during inflation may be the more likely possibility, and,
perhaps disappointingly, it may
have little consequence for cosmological or astrophysical observations.

\noindent
{\bf Acknowledgements:}
We thank T. Banks, D. Lyth, S. Thomas and P.J. Fox for conversations and 
criticism. A. Anisimov thanks M.Trusov for helping with numerical simulations
at the early stage of this work.
\noindent
This work supported in part by the U.S.
Department of Energy and by RFBR grant
02-02-17379.


\end{document}